\documentclass{elsart}

\usepackage[T1]{fontenc}
\usepackage[latin1]{inputenc}
\usepackage{graphics}
\usepackage{verbatim}
\usepackage{amsmath}

\begin{document} 
\begin{frontmatter} 
\newcommand{\bd}[1]{ \mbox{\boldmath $#1$}  }
\newcommand{\xslash}[1]{\overlay{#1}{/}}
\newcommand{\sla}[1]{\xslash{#1}}
\newcommand{\qn}{\mathbf{q}}
\newcommand{\rn}{\mathbf{r}}
\newcommand{\Rn}{\mathbf{R}}


\title{Uncorrelated scattering approximation revisited}

\author[tagus,famn]{A. M. Moro\corauthref{cor1}},
\ead{moro@nucle.us.es}
\corauth[cor1]{Corresponding author}
\author[famn]{J. A. Caballero},
\ead{juan@nucle.us.es}
\author[famn]{J. G\'omez-Camacho}
\ead{gomez@nucle.us.es}
\address[tagus]{Departamento de
F\'{\i}sica, Instituto Superior T\'ecnico, Taguspark,
Av. Prof. Cavaco e Silva,  Taguspark,
2780-990 Porto Salvo, Oeiras, Portugal}
\address[famn]{Departamento de F\'{\i}sica At\'omica, Molecular y Nuclear, 
Universidad de Sevilla,  
Apdo. 1065, E-41080 Sevilla, Spain} 
 
\vspace{0.5cm} 

\begin{abstract}
The formalism to describe the scattering of a weakly bound projectile nucleus 
by a heavy target is investigated, using the Uncorrelated 
Scattering Approximation. The main assumption involved is to 
neglect the 
correlation between the fragments of the projectile in the region where the 
interaction with the target is important. It is shown that the angular
momentum of each fragment with respect to the target is conserved. Moreover,
when suitable approximations are assumed, the kinetic energy of each fragment
is also shown to be conserved. The $S$-matrix for the scattering of
the composite system can be written as a combination of terms, each one
being proportional to the product of the $S$-matrices of the fragments. 
\end{abstract}

\end{frontmatter}




\section{Introduction\label{section:intro}}


In recent years there has been much interest in the description of halo 
nuclei. These are weakly bound and spatially extended nuclear systems, so that
some fragments of the nucleus (generally neutrons) have a high probability
of being at distances larger than the typical nuclear radii 
(see Refs.~\cite{Han95,Zhu93} for a general review on these nuclei).

Reactions induced by halo nuclei on different targets are a useful
tool to investigate the structure of these nuclei. A proper description of the
reaction mechanism of the collision of a halo nucleus with a target is a
complex quantum mechanical problem, which requires the explicit treatment
of the continuum of breakup states. This treatment can be performed by 
discretizing the continuum of breakup states. However, the price to be paid 
is that the relation between the scattering observables measured  
and the structural properties involved in the problem becomes
troublesome.

When the scattering energy of the projectile is sufficiently high,
the interaction of the halo nucleus with the target can be considered to occur
in such a short time that the fragments of projectile are practically
``frozen''  in given relative positions.
This ``frozen halo'', also known as  ``adiabatic'' 
approximation, is the basis of different approximate treatments which have 
been successfully applied to the interpretation of 
halo nuclei scattering 
measurements \cite{Glau59,Ron97,Ron97b}.

The validity of the ``frozen halo'' approach depends on the comparison of
the collision time $\tau_C$ and the internal time $\tau_I$. The collision 
time can be estimated as $\tau_C = a/v$, where $a$ is a typical length scale
of the interaction and $v$ is the relative velocity. The internal time
is given by $\tau_I = \hbar/\epsilon$ where $\epsilon$ is a typical 
excitation energy of the projectile. Following this argument, the 
``frozen halo'' approximation will be valid provided that the collision time
is much shorter than the internal time $\tau_C \ll \tau_I$. 

Nevertheless, tidal forces which tend to
distort the structure of the projectile while interacting with the target, are also
very important. This distortion has a time 
scale $\tau_D$ which can be estimated in terms of the tidal force
$F_T = V_T/a$ through the expression $a = 1/2 F_T/m \tau_D^2$. Thus, the 
validity of the ``frozen halo'' approximation requires also
$\tau_C \ll \tau_D$. 

In this paper we focus on a different approach named the 
Uncorrelated Scattering Approximation (USA), that we first introduced in
\cite{Mor01a}.
Its basic approximation consists in assuming that 
the interaction between the fragments of the halo nuclei can be neglected
during the collision time. Thus, the fragments of 
the projectile scatter separately from the target, each of them with a fraction of the
total energy of the projectile.
The validity of the USA is linked to a correlation time 
$\tau_R = \hbar /\Delta$, where $\Delta$ measures the energy spread of the
ground state of the projectile when the interaction between the fragments 
is neglected. So, the USA is expected to be valid when $\tau_C \ll \tau_R$,
but, in contrast with the ``frozen halo'' approach, it does not depend on the
strength of the tidal forces.

The adiabatic or ``frozen halo'' approximation implies that the relative 
coordinates within the projectile are constant during the scattering process.
As we shall show, the USA implies that the relative angular 
momentum between each fragment of the projectile and the target is conserved.
Moreover, the kinetic energy of each fragment is also conserved.

In section 2 we develop the formalism of the uncorrelated scattering 
approximation (USA) making use of the expansion of the
three-body $T$-matrix. 
In section 3 we present an application of the USA approach
to the case of elastic scattering and breakup of deuteron  on 
\nuc{58}{Ni}. In section 4 the conclusions are presented.


\section{The Uncorrelated Scattering Approximation from a $T$-matrix approach}

We consider a two-body system ($A,B$) which is scattered 
from a heavy target ($T$).
Neglecting spin and other internal degrees of freedom of the fragments and
the target, the Hamiltonian is given by
\begin{eqnarray}
\label{eq:Hmod}
H&=&\frac{\vec{P}^{2}}{2M}+\frac{\vec{p}^{2}}{2\mu}+v_{AB}
+v_{AT}+v_{BT} \nonumber \\
 &=&\frac{\vec{P}_{A}^{2}}{2m_A}+\frac{\vec{P}_{B}^{2}}{2m_B}+v_{AB}
+v_{AT}+v_{BT} \, ,
\end{eqnarray}
where $M=m_A+m_B$ and $\mu=m_A m_B/(m_A+m_B)$. The interactions $v_{AB}$, 
$v_{AT}$ and  $v_{BT}$ depend on the relative coordinate and momenta of the 
interacting particles. 

To define a scattering process, we consider an asymptotic situation in which
the fragments are very far from the target. In this case, the asymptotic
Hamiltonian is given by
\begin{equation}
\label{eq:H0}
H_0= \frac{\vec{P}^{2}}{2M}+\frac{\vec{p}^{2}}{2\mu}+v_{AB} = 
\frac{\vec{P}^{2}}{2M}+h \, .
\end{equation}

The eigenstates of $H_0$ corresponding to a given total energy $E$ 
which are relevant for a scattering process can be
expressed as a free wave $|k_i; LM_L\rangle$ on the relative
coordinate times an eigenstate $|\phi_i; I M_I\rangle$ of the internal
Hamiltonian $h$. Note that these states can be characterized by a given
angular momentum and projection.
It should be noticed that the eigenstate $|\phi_i; I M_I\rangle$
corresponding to an eigenvalue $\epsilon_i$ of $h$,
may be a bound or a continuum state. The momentum $k_i$ is
related to the energy of the state through 
$\hbar^2 k_i^2/ 2 M + \epsilon_i = E$. 
We can couple the product states to a total angular momentum $J$, to give
$|\phi_i,k_i;L,I,JM_J\rangle$. 
Since spin projection $M_J$ is conserved,  it will be dropped in the following 
derivation.
The effect of the interaction $v_{AT}+v_{BT}$ of the projectile
with the target is given by a $T$-matrix, connecting the eigenstates of 
$H_0$ with energy $E$. 
Formally, the $T$-matrix can be expressed as
\begin{equation}
T(E) = v_{AT} + v_{BT} + (v_{AT} + v_{BT}) {1 \over E^+ - H_0} T(E) \, ,
\end{equation}
where $E^+ =E+i\epsilon$.

The scattering amplitude is proportional to the on-shell matrix elements
connecting the initial bound state $|\phi_g,k_g;L,I,J\rangle$ 
with a final state $|\phi_f,k_f;L',I',J\rangle$, which may be bound 
or unbound. 
Here, the quantities $k_g$ and $k_f$ represent the relative momenta of the
two-body system ($A,B$) with respect to the target $T$ in the initial and
final channels, respectively.
The $S$-matrix can be expressed as
\begin{eqnarray}
 && \langle\phi_f,k_f;L',I',J|S|\phi_g,k_g;L,I,J\rangle = 
\nonumber \\ &&
=\delta_{f,g}  + 2 \pi i \langle\phi_f,k_f;L',I',J|T(E)|\phi_g,k_g;L,I,J\rangle \, .
\label{eq:SvsT}
\end{eqnarray}
Then, in order to apply the Uncorrelated Scattering Approximation, we make use of the
fact that the interaction $v_{AT}+v_{BT}$ conserves separately
the angular momenta $L_A$ and $L_B$ of the fragments of the projectile with
respect to the target. Note that the $T$-matrix will not, in general, conserve
$L_A$ and $L_B$, due to the interaction $v_{AB}$ between the fragments, which
is included in the propagator. However, a reasonable approach is to
neglect the term $v_{AB}$ in the propagator, while we rescale it 
by a factor $\lambda$, which can be a function of $J$, and will be determined 
later. This approximation implies dismissing the relatively weak interaction
$v_{AB}$ in the region where $v_{AT}+v_{BT}$
is strong (it contributes to second order to the $T$-matrix). 
Hence, we make the approximation
\begin{equation}
{1 \over E+i\epsilon - H_0} \simeq  {\lambda \over E+i\epsilon - \bar H_0} \, ,
\end{equation}
where $\bar{H}_0 = K_A + K_B$ is the Hamiltonian which contains only the 
kinetic energy terms.
The factor $\lambda$ included in this expression may be complex, describing in that
case the loss of flux due to the excitation of the target and/or the
fragments of the projectile.

The above approximation leads to the expression
\begin{eqnarray}
T(E) &\simeq& {1\over \lambda} \bar T(E) \\
\bar T(E) &=& \lambda(v_{AT} + v_{BT}) + \lambda(v_{AT} + v_{BT}) 
{1 \over E^+ - \bar H_0} \bar T(E) \, .
\end{eqnarray}
Note that $\bar T(E)$ is the $T$-matrix for the 
scattering of two uncorrelated particles $A$ and $B$ with a renormalized 
interaction $\lambda (v_{AT} + v_{BT})$.

The eigenstates $|\phi_i,k_i;L,I,J\rangle$ 
of $H_0$ can be expanded in terms of eigenstates of $\bar H_0$,
\begin{equation}
|\phi_i,k_i;L,I,J\rangle = \int d\alpha g_i(\alpha)
|\alpha,E_i(\alpha);L,I,J\rangle \, .
\end{equation}
In the above expressions, $E_i(\alpha)=(E-\epsilon_i)\sec^2(\alpha)$ is the 
eigenvalue of $\bar H_0$  with 
$\alpha$ the hyperangle which depends on the ratio of the internal 
momentum $q$ and the relative momentum $k_i$, i.e.,
$q(k_i,\alpha) = k_i \sqrt{\mu/M} \tan{\alpha}$.
The explicit expression of $g_i(\alpha)$ is given in the 
appendix \ref{sec:ap-B}.
 
The matrix elements of $\bar T(E)$ become
\begin{eqnarray}
&& \langle\phi_f,k_f;L',I',J|\bar T(E)|\phi_g,k_g;L,I,J\rangle \nonumber \\
&=&\int \int d\alpha' d\alpha g_f(\alpha')  g_g(\alpha)  
\langle\alpha',E_f(\alpha');L',I',J|\bar T(E)|\alpha,E_g(\alpha);L,I,J\rangle \, .
\label{talfa}
\end{eqnarray}
Thus, Eq.~(\ref{talfa}) reflects that the on-shell matrix element of $\bar T$, 
between eigenstates of $H_0$ corresponding to the energy $E$, are
given in terms of
off--shell matrix elements of $\bar T$ between eigenstates of $\bar H_0$
corresponding to different energies $E_{g}(\alpha)$, $E_{f}(\alpha')$.

Besides the assumptions involved in (\ref{talfa}), in what follows we
make a further approximation by replacing the
off--shell matrix elements of $\bar T(E)$ by on-shell matrix elements. So,
while we retain the $\alpha$ dependence of the matrix elements, we
substitute the energy dependence for the on-shell value. Thus, 
the matrix elements in (\ref{talfa}) can be approximated by
\begin{eqnarray}
&& \langle\alpha',E_f(\alpha');L',I',J|\bar T(E)|\alpha,E_g(\alpha);L,I,J\rangle
\nonumber \\
&\simeq & \lambda A^f(\alpha')
\langle\alpha',E;L',I',J|\bar T(E)|\alpha,E;L,I,J\rangle
A^g(\alpha) \, . 
\label{talfaonshell}
\end{eqnarray}
The factors $A^f(\alpha')$ and $A^g(\alpha)$ are introduced to
ensure unitarity.
Note that the typical values of $E_{i}(\alpha)-E$ are of the order of the
kinetic energy of the relative motion of the fragments, which is related to the
correlation time through $\hbar / \tau_R$. The operator 
$\bar T(E)$ connects states in a range of energies which is determined by 
the collision time through $\Delta E = \hbar/\tau_C$. 
So, this on-shell approximation is justified provided that the correlation 
time is long compared to the collision time.

Now, for the purpose of evaluating the on-shell matrix elements of $\bar T(E)$
we can make use of an expansion in hyperspherical harmonics, which allows us
to write down the operator $\bar T(E)$ between states with definite values
of the angular momenta $L_A$ and $L_B$ of each fragment with respect to the 
target. Note that these magnitudes are conserved by the interaction. After some
algebra (see Appendix \ref{sec:ap-A} for details), we finally get
\begin{eqnarray}
&&\langle \phi_f,k_f;L',I',J|T(E)|\phi_g,k_g;L,I,J\rangle 
\nonumber \\
&\simeq &\sum_{KK'}\sum_{L_A,L_B}
\langle\phi_f|K'\rangle_{L'I'}\langle K|\phi_g\rangle_{LI}
\langle L_A,L_B|L,I\rangle_{KJ}\langle L',I'|L_A,L_B\rangle_{K'J} \nonumber
\\
&\times & 
\langle K',E;L_A,L_B,J|\bar T(E)|K,E;L_A,L_B,J\rangle \label{a1}
\end{eqnarray}
with
\begin{eqnarray}
&&\langle K',E;L_A,L_B,J|\bar T(E)|K,E;L_A,L_B,J\rangle \,\, \nonumber \\
& =& \int\int d\beta d\beta'
f_{L_A,L_B}^K(\beta) f_{L_A,L_B}^{K'}(\beta')
\langle\beta',E;L_A,L_B,J|\bar T(E)|\beta,E;L_A,L_B,J\rangle, \nonumber \\
\label{a2}
\end{eqnarray}
where the functions $f_{L_A,L_B}^K(\beta)$  and 
the overlaps $\langle K|\phi_g\rangle_{LI}$ are given in the appendix
\ref{sec:ap-A}, and
$\langle L_A,L_B|L,I\rangle_{KJ}$ are the Raynal-Revay coefficients 
\cite{Ray70}. The hyperangular variable $\beta$ determines the partition
of the energy $E$ between the fragments $A$ and $B$ of the projectile, so that
$e_A=E \cos^2(\beta)$ and $e_B=E \sin^2(\beta)$.  

Expanding the $T$-matrix up to third order and using the results of
appendix \ref{sec:ap-C} the matrix elements of the operator $\bar T(E)$
can be written as:
\begin{eqnarray}
\langle\beta',E| \bar T(E) |\beta,E\rangle
=  \delta(\beta-\beta') 
\Big\{  t_A(e_A) + t_B(e_B) + 2 i \pi  t_A(e_A)t_B(e_B) \Big\}\, ,
\end{eqnarray}
where $t_A(e_A)$ denotes a on-shell two-body matrix element for the
$A-T$ scattering (analogously for $t_B$).


Recovering the angular momenta, the matrix elements
of the $T$-matrix between states with given $K,L_A,L_B,J$ values, result 
\begin{eqnarray}
&&\langle K',E;L_A,L_B,J|\bar T(E)|K,E;L_A,L_B,J\rangle =
\int d \beta f^{K'}_{L_A,L_B}(\beta)f^{K}_{L_A,L_B}(\beta) \nonumber \\
&\times & \Big\{t_A(E \cos^2\beta)+ t_B(E \sin^2\beta) + 
2 \pi i t_A(E \cos^2\beta) t_B(E\sin^2\beta)\Big\}
\, ,
\end{eqnarray}
which, according to Eq.~(\ref{eq:SvsT}), can be written
in terms of  $\bar S$,  i.e., the
$S$-matrix which describes the scattering
with boundary conditions given by $\bar H_0$, as
\begin{eqnarray}
&&\langle K',E;L_A,L_B,J|\bar S|K,E;L_A,L_B,J\rangle 
\nonumber \\
&=& \int d \beta 
f^{K}_{L_A,L_B}(\beta)f^{K'}_{L_A,L_B}(\beta)
S_A(E \cos^2\beta) S_B(E \sin^2\beta) \, .
\end{eqnarray}

If the values of $K,K'$
are restricted by a maximum hyperangular momentum $K_m$,
the above integral can be approximated by quadratures.  It 
is particularly convenient to consider
the quadrature associated to the Jacobi polynomials that determine the
functions $f^{K}_{L_A,L_B}(\beta)$ with
a number of points $N$ equal to the number of $K$ values allowed.
Explicitly:
\begin{eqnarray}
& & \langle K',E;L_A,L_B,J|\bar S|K,E;L_A,L_B,J\rangle 
\nonumber \\ 
&=& 
\sum_{n=1}^N \langle K'|n\rangle_{L_A,L_B}
\langle n|K\rangle_{L_A,L_B} 
S_A(e_{A}^{n}) S_B(e_{B}^{n}) \, ,
\end{eqnarray}
where, following \cite{Mor01a,tesismoro},  
the coefficients $\langle K|n\rangle_{L_A,L_B}$ 
are given by
\begin{equation}
\langle K|n\rangle_{L_A,L_B} = 
\frac{\sqrt{w_n}}
{(\cos\beta_n)^{L_A+1}(\sin\beta_n)^{L_B+1}}
f^{K}_{L_A,L_B}(\beta_n)  \, .
\end{equation}
The quadrature points $\beta_n$ correspond to the zeros
of the function $f^{K_m}_{L_A,L_B}(\beta)$. The explicit expression
for the quadrature weights ${w_n}$ can be found in \cite{Mor01a}.
Note that, in terms of the quadrature points, the individual energies
of particles $A$ and $B$ are given by
$e_{A}^{n}=E \cos^2(\beta_n)$ and $e_{B}^{n}=E \sin^2(\beta_n)$.


Then, collecting  these results, we may write
\begin{eqnarray}
&& \langle\phi_f,k_f;L',I',J| S(E)|\phi_g,k_g;L,I,J\rangle
  \nonumber \\  
& \simeq &\sum_{L_A,L_B,n} 
\langle\phi_f,k_f;L',I',J|n; L_A, L_B, J\rangle
S_A(e_{A}^{n}) S_B(e_{B}^{n})
\langle n; L_A, L_B, J|\phi_g,k_g;L,I,J\rangle , \nonumber \\
&& \label{norts}
\end{eqnarray}
with
\begin{equation}
\langle \phi_i,k_i;L,I,J|n; L_A, L_B,J\rangle \equiv 
\sum_{K} \langle\phi_i|K\rangle_{L,I}
\langle L,I|L_A,L_B\rangle_{KJ} \langle K|n\rangle_{L_A,L_B}. \,\,\,\,\,
\end{equation}

To finish with this analysis we proceed by 
evaluating the parameter $\lambda$ which 
determines the renormalization of the interactions $v_A$ and $v_B$. 
For this purpose, we write the individual $S$-matrices in terms of the phase shifts
\begin{equation}
S_A(e_{A}^{n}) = \exp\Big(2 i \delta(\lambda v_A;L_A,e_{A}^{n})\Big) 
\,; \,\,\,
 \,\,\,\,\,\,
S_B(e_{B}^{n}) = \exp\Big(2 i \delta(\lambda v_B;L_B,e_{B}^{n})\Big) \, .
\label{eq:sasb}
\end{equation}
In many approximations the phase shift is found to be  proportional to the potential. 
Under this situation, the phase shift
scales with the parameter $\lambda$. This is for instance
the case of the eikonal and WKB approximations, where the phase shift is obtained
as an integral of the potential along a classical trajectory. In general, there is not
a simple relation between the phase shift
and the potential, but only for the purpose of estimating in a simple
way the value of $\lambda$, we will assume that this relation holds approximately. So, $\delta(\lambda v_A;L_A,e_{A}^{n}) = \lambda \delta(v_A;L_A,e_{A}^{n})$.



In the limit of weak interactions, one expects that breakup effects are
negligible, and thus the elastic $S$ matrix should be well described by the
folding interaction. Let us assume that the ground state has $I=0$, implying $L=J$.
Then, the folding interaction, which depends on the relative coordinate
and momentum is given as
\begin{equation}
\label{eq:vfold}
v_F=\langle \phi_g|v_{AT}+v_{BT}|\phi_g \rangle
\end{equation}
and the expression for the elastic $S$-matrix due to the
folding interaction can be written as
\begin{equation}
S_F(E,J) = \exp\Big(2 i \delta(v_F;J,E)\Big) \, .
\label{eq:sfold}
\end{equation}
Thus, comparing  Eqs.~(\ref{eq:sfold}) and (\ref{norts})
in the limit of weak interactions, 
one gets
\begin{equation}
 \delta(v_F;J,E) = \lambda \bar{\delta}_{LIJ}\, ,
\label{eq:lambda}
\end{equation}
where
\begin{eqnarray}
\bar{\delta}_{LIJ} &=&
\sum_{n, L_A, L_B} |\langle n; L_A, L_B, J|\phi_g,k_g;L,I,J\rangle|^2 
\Big\{\delta(v_A;L_A,e_A)+\delta(v_B;L_B,e_B)\Big\} \nonumber \, .\\
&&
\end{eqnarray}

Using this result, the product of $S$-matrices can be written as
\begin{equation}
S_A(e_{A}^{n})S_B(e_{B}^{n}) = 
\exp\Big(2 i \pi p(n,L_A,L_B,J) \delta(v_F;J,E)\Big) 
\label{power} \, ,
\end{equation}
where
\begin{equation}
p(n,L_A,L_B,J)=
\Big\{
\delta(v_A;L_A,e_{A}^{n})+\delta(v_B;L_B,e_{B}^{n})
\Big\}/
\bar{\delta}_{LIJ}.
 \end{equation}
Note that $p(n,L_A,L_B,J)$ is a real, positive number, 
with typical values around unity. It fulfils the relation
\begin{eqnarray}
 \sum_{n, L_A, L_B} p(n,L_A,L_B,J) 
|\langle n; L_A, L_B, J|\phi_g,k_g;L,I,J\rangle|^2 &=& \nonumber \\
= \sum_{n, L_A, L_B}|\langle n; L_A, L_B, J|\phi_g,k_g;L,I,J\rangle|^2 
&=& 1
\, \, \,. 
\end{eqnarray}
The value of $p(n,L_A,L_B,J)$ measures how strong is the interaction of
the fragments with the target in the configuration which is characterized
by the quantum numbers $n,L_A,L_B$, compared with the average interaction
over all the configurations that contribute to the elastic channel.
Note that Eq.~(\ref{power}) implicitly states that any absorption effect,
due to the excitation of the target or the fragments of the projectile,
that can be described by the use of a complex folding potential, 
will also appear in the uncorrelated three-body calculation, 
scaled by the factor $p(n,L_A,L_B,J)$.

We would like to stress that Eq.~(\ref{norts}) can be 
applied to elastic, inelastic and breakup 
scattering. In particular, the elastic $S$-matrix elements are given by
\begin{eqnarray}
&& \langle\phi_g,k_g;L,I,J| S(E)|\phi_g,k_g;L,I,J\rangle
  = \nonumber  \\
&& \sum_{L_A,L_B,n} |\langle n; L_A, L_B, J|\phi_g,k_g;L,I,J\rangle|^2 
S_A(e_{A}^{n}) S_B(e_{B}^{n}) \, . 
\label{eq:Selastic}
\end{eqnarray}
The inelastic $S$-matrix to an excited bound state $|\phi_e\rangle$ is
\begin{eqnarray}
&& \langle\phi_e,k_e;L',I',J| S(E)|\phi_g,k_g;L,I,J\rangle
  = \nonumber  \\
&& \sum_{L_A,L_B,n} \langle n; L_A, L_B, J|\phi_g,k_g;L,I,J\rangle 
\langle \phi_e,k_e;L',I',J|n; L_A, L_B, J\rangle 
S_A(e_{A}^{n}) S_B(e_{B}^{n}) \, . \nonumber \\ 
\label{eq:Sinelastic}
\end{eqnarray}
%
It is also possible to calculate the  $S$-matrix elements 
for the breakup leading to specific states
of the continuum, characterized by an asymptotic linear
internal momentum $p$ and quantum numbers \{$L'$,$I'$,$J$\}. According to 
Eq.~(\ref{norts}) these are given by
\begin{eqnarray}
S^{J}_{LI;L'I'}(p)&\equiv &
 \langle\phi(p),k(p);L',I',J| S(E)|\phi_g,k_g;L,I,J\rangle
 \nonumber  \\
&=& \sum_{L_A,L_B,n} 
\langle\phi(p),k(p);L',I',J| n; L_A, L_B, J\rangle 
\nonumber \\
&\times&\langle n; L_A, L_B, J|\phi_g,k_g;L,I,J\rangle 
S_A(e_{A}^{n}) S_B(e_{B}^{n}) \, .
\label{eq:Sbreakup}
\end{eqnarray}
Note that, due to energy conservation, $\hbar^2 k^2(p)/2M+\hbar^2p^2/2\mu=E$.
The differential breakup cross section is calculated directly from
the above $S$-matrix elements by means of the 
expression 
\begin{equation}
\frac{d\sigma ^{J}_{LI;L'I'}(p)}{dp}=\frac{\pi }{k^{2}_{g}}
\frac{2J+1}{2I+1}\left| S_{LI;L'I'}^{J}(p)\right| ^{2} .
\label{eq:difbu}
\end{equation}

An appealing feature of our previous formulation of the USA,  
presented in \cite{Mor01a}, is that 
it leads to compact expressions for the integrated breakup cross 
sections. Although the derivation presented in this work differs
to some extent from our previous formulation, it also leads to similar 
closed formulae for the breakup cross sections. In particular, 
the total integrated breakup cross section to all the final continuum
states characterized by the angular momenta ${L',I',J}$ 
is given by  (c.f.~ with Eq.~(36) of Ref.~\cite{Mor01a}): 

\begin{eqnarray}
\sigma_J ^{bu}(LI\to L'I') & = & \frac{\pi }{k^{2}_{g}}\frac{2L+1}{2I+1}
\Big {\{}\sum _{K=L'+I'}^{K_m}
|\langle K,L',I',J|\bar S|\phi_g,k_g; L,I,J\rangle |^{2}\nonumber \\ 
& -&  \sum_e
|\langle \phi_e,k_e;L',I',J|\bar S|\phi_g,k_g;L,I,J\rangle |^{2} 
\nonumber \\ 
&-& \delta_{L,L'}\delta_{I,I'}
|\langle \phi_g,k_g;L,I,J|\bar S|\phi_g,k_g;L,I,J\rangle |^{2}\Big {\}}.
\label{eq:xsecbuLIJ} 
\end{eqnarray}
This expression is readily obtained by summing upon $L'$ and $I'$ in
Eq.~(\ref{eq:difbu}) and applying the 
closure relation (\ref{closure}) to the final states.


It is also possible to obtain a compact expression for the 
breakup cross section
corresponding to a total angular momentum $J$, $\sigma ^{bu}_{J}$.
This is achieved upon summation of $\sigma_J ^{bu}(LI\to L'I')$ 
on the angular
momenta $L'$ and $I'$ and taking into account the completeness property
of the states $|K, L',I',JM_{J}\rangle $. This leads to the closed expression
\begin{eqnarray}
\label{eq:xsecbuJ}
\sigma ^{bu}_{JLI} &= &\frac{\pi }{k^{2}_{g}}\frac{2L+1}{2I+1}
\Big{\{}\langle \phi_g,k_g;L,I,J|\bar S^{+} \bar S|\phi_g,k_g;L,I,J\rangle  
\nonumber \\
&-&  \sum_{e,L',I'}
|\langle \phi_e,k_e; L',I',J|\bar S|\phi_g,k_g;L,I,J\rangle |^{2}  
\nonumber \\
&-&
|\langle \phi_g,k_g; L,I,J|\bar S|\phi_g,k_g;L,I,J\rangle |^{2}\Big{\}}. 
\end{eqnarray}
\indent Then, within the USA, the integrated breakup cross section for a given total 
angular momentum is calculated as the dispersion of the operator 
$\bar S$ in the ground state of the projectile, subtracting the 
contribution of the other bound states.

In summary, in this section we have derived an approximate formula to evaluate
the elastic, inelastic and breakup $S$-matrix elements corresponding to a three-body scattering 
process. These approximated 
$S$-matrix elements are expressed in compact form by Eq.~(\ref{norts}). The 
simplicity
of this formula relies on the fact that it relates the complicated three-body
collision matrix to a simple superposition of  two--body $S$-matrices, 
weighted by some factors
which depend on  analytical coefficients and the 
structure of the composite. Expression (\ref{norts}) provides also a possible 
physical interpretation for the states $|n;L_A,L_B, J\rangle$. These 
states are eigenstates
of the Hamiltonian $\bar{H}$ in a discrete basis. According to 
(\ref{norts}), the initial 
state is decomposed in the full set of states $|n;L_A,L_B, J\rangle$. Upon neglection 
of the inter-cluster interaction
$v_{AB}$, the two particles evolve in this set of states, 
interacting with the target through
the interactions $v_A$ and $v_B$, but preserving the individual 
energies and angular momenta of the constituents. The 
superposition of states resulting after 
the interaction with the target is finally
projected onto the physical final state for the process under consideration,
denoted $|\phi_f,k_f;L',I',J\rangle$.

\section{Application to deuteron scattering on \nuc{58}{Ni}}
As an illustrative example, in this section we apply the USA method
to the reaction d+\nuc{58}{Ni} at $E_d$=80~MeV. This reaction has been
previously analyzed  by several authors within the CDCC scheme 
\cite{Aus87,Piy99,Ise86,Yah86}, providing
a satisfactory agreement with the 
existing elastic data \cite{Ste83}. Therefore,
in this work we have adopted  the CDCC calculation as 
the benchmark result
to compare our method with. In order to obtain the  elastic and
breakup observables, we performed  CDCC calculations similar to 
those reported in \cite{Aus87,Piy99}.
The proton-target and neutron-target interactions were described in
terms of the Becchetti-Greenless global parameterization \cite{Bec69}.
The p-n interaction, which is required to construct the deuteron
ground state and continuum bins, was parametrized as 
$v(r)=v_{0}\exp(-r^{2}/a^{2})$,
with $v_{0}=-72.15$~MeV and $a=1.484$~fm \cite{Aus87}. Following 
the standard procedure, the p-n continuum was discretized
into energy bins. The partial waves $I$=0, 2, 4, 6 and 8 were included in the
CDCC model space. The odd partial 
waves were not considered, since their 
influence on the dynamics turned out to
be  negligible. The maximum excitation energies for these waves
were $\epsilon_{\max}$=40 MeV ($s$-waves), 45 MeV ($d$ and $f$-waves), 
50 MeV ($g$-waves) and 60 MeV ($h$-waves).
For simplicity,
the proton and neutron spins are ignored. These CDCC calculations
were carried out with the computer code FRESCO \cite{Thom88}.

The USA calculations were performed using Eq.~(\ref{norts}). According 
to this 
expression, the three-body $S$-matrix  is constructed
by superposition of the on-shell cluster--target $S$-matrices, evaluated
at the appropriate energies and angular momenta. For computational convenience, we
found useful to evaluate these $S$-matrices using the 
WKB approximation. This has the extra
advantage that it makes exact the scaling property of the phase shift with the 
parameter $\lambda$, thus making more sensible the estimate (\ref{eq:lambda})
for this parameter.


\subsection{Elastic scattering}
Since the USA method is  formulated in terms of the $S$-matrix,
we first compare the elastic $S$-matrix elements in both approaches. 
The ground state wavefunction was described with the same Gaussian 
potential used for the CDCC calculation.



\begin{figure}
{\centering \resizebox*{0.75\columnwidth}{!}
{\includegraphics{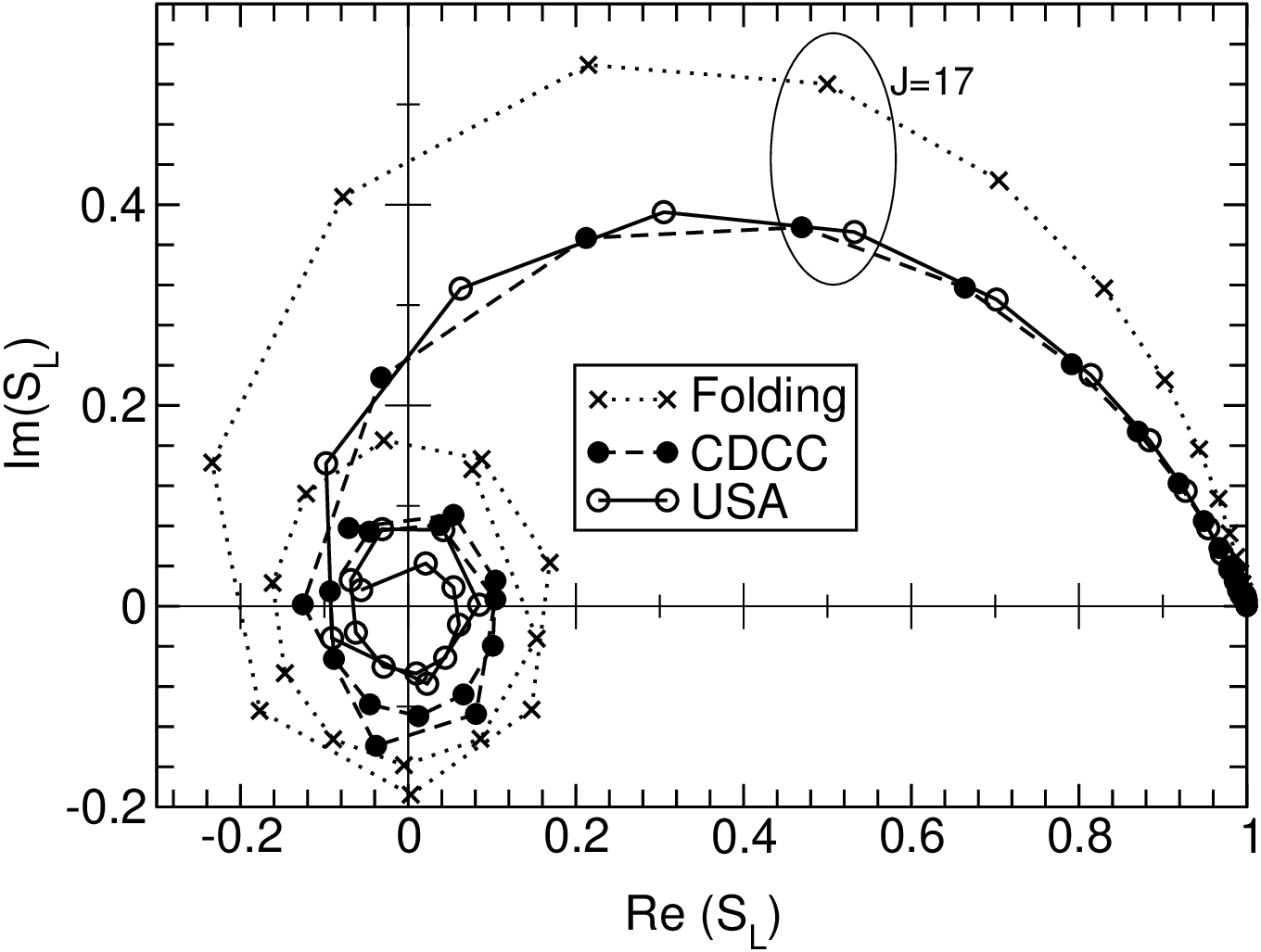}} \par}
\caption{\label{Fig:argand}Argand plot for the CDCC (filled circles), USA
(open circles) and cluster-folded (crosses) calculations, corresponding
to the d+$^{58}$Ni elastic scattering at $E_d$=80~MeV.}
\end{figure}

In Fig.~\ref{Fig:argand} we present the CDCC and USA calculations
in the form of an Argand plot, in which the real and imaginary parts
of the $S$-matrix elements are represented in the $x$ and
$y$ axis, respectively. Note that, since the internal spin is
neglected, the orbital angular momentum for the projectile-target
relative motion $L$ coincides with the total angular momentum
$J$. The CDCC and USA calculations are represented by filled and open circles,
respectively. 
For comparison purposes, we have also included the cluster-folded
(CF) calculation (crosses), in which the projectile-target interaction is given
by the sum of the proton-target and neutron-target interactions, folded
with the deuteron density, according to Eq.~(\ref{eq:vfold}). Note 
that the latter is equivalent to a
CDCC calculation without continuum bins. Therefore, the difference between
the folding and the CDCC calculations evidences the effect of the breakup
channels on the elastic scattering. From the curves in Fig.~\ref{Fig:argand},
it becomes apparent that the presence of the continuum  reduces
the modulus of the $S$-matrix for all the partial waves, due
to the loss of flux to the breakup channels.  It can be 
seen that the USA method also succeeds 
on reproducing this deviation
from the CF calculation, giving a result close to the CDCC.

\begin{figure}
{\centering \resizebox*{0.75\columnwidth}{!}
{\includegraphics{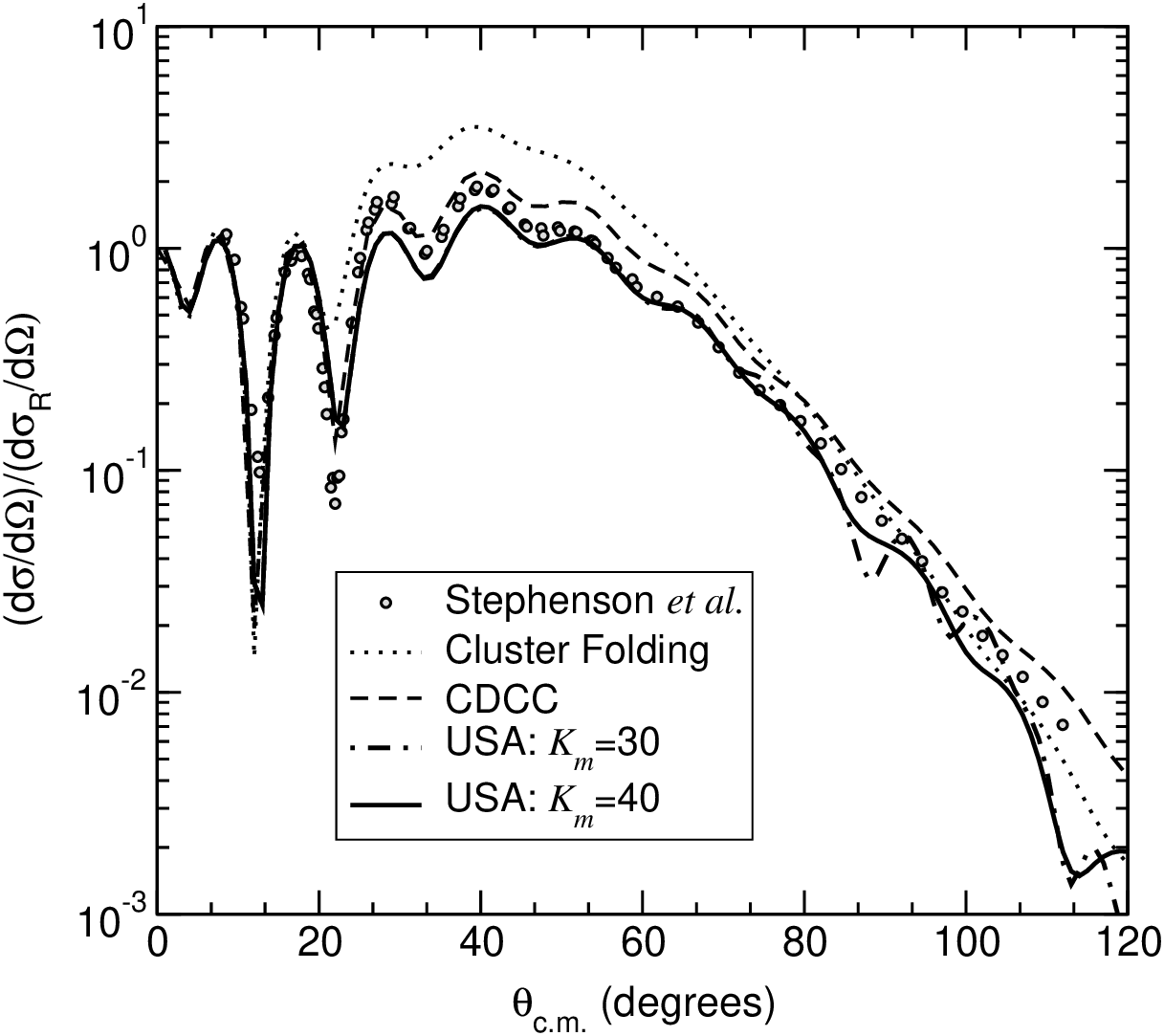}} \par}
\caption{\label{Fig:dni-el}
Elastic scattering angular distribution for the
d+$^{58}$Ni reaction at $E_{d}$=80~MeV.
The meaning of the curves are indicated by the labels.}
\end{figure}

The differential elastic angular distributions derived from 
the above $S$-matrices are
shown in Fig.~\ref{Fig:dni-el}, along with the data of 
Stephenson {\it et al.} \cite{Ste83}.
The dotted line represents the folding calculation, which clearly overestimates
the data, particularly at intermediate angles. Inclusion of the continuum
within the CDCC scheme (dashed line), produces a significant reduction
of the elastic cross section and improves significantly the agreement
with the data. Notice that, despite this improvement, the data are still
somewhat overestimated at angles around 60$^\circ$.
Two different USA calculations are presented  in Fig.~\ref{Fig:dni-el}.
The first one, represented in the graph by the dotted--dashed line, was 
performed  with a maximum hyperangular
momentum $K_{m}$=30. This calculation agrees fairly well with
the data in the whole angular range. Note that for 
$\theta_\mathrm{c.m.}<25^\circ$ the 
agreement with the data is excellent. At intermediate angles, 
the USA reproduces  well the interference pattern, although 
the absolute normalization is somewhat underestimated.
This suggests that the method will overestimate the breakup cross
sections, as it will be discussed below. For c.m.
angles above 80$^\circ$ the USA calculation
shows some oscillations, which are not present on neither
the CDCC calculation nor the experimental data. These spurious wiggles
are partially suppressed when the  cutoff hyperangular momentum is increased
to $K_{m}$=40, as shown by the solid line. We found that for 
higher values of $K_{m}$ the angular
distribution remains basically unchanged. Thus, we can conclude that
for $K_{m}$=40 we have convergence of the USA calculation.

\subsection{Breakup reaction}
We next analyse the breakup channel for the same reaction. Due 
to the absence of experimental data, we will compare
our model directly with CDCC. 

\begin{figure}
{\centering \resizebox*{0.75\columnwidth}{!}
{\includegraphics{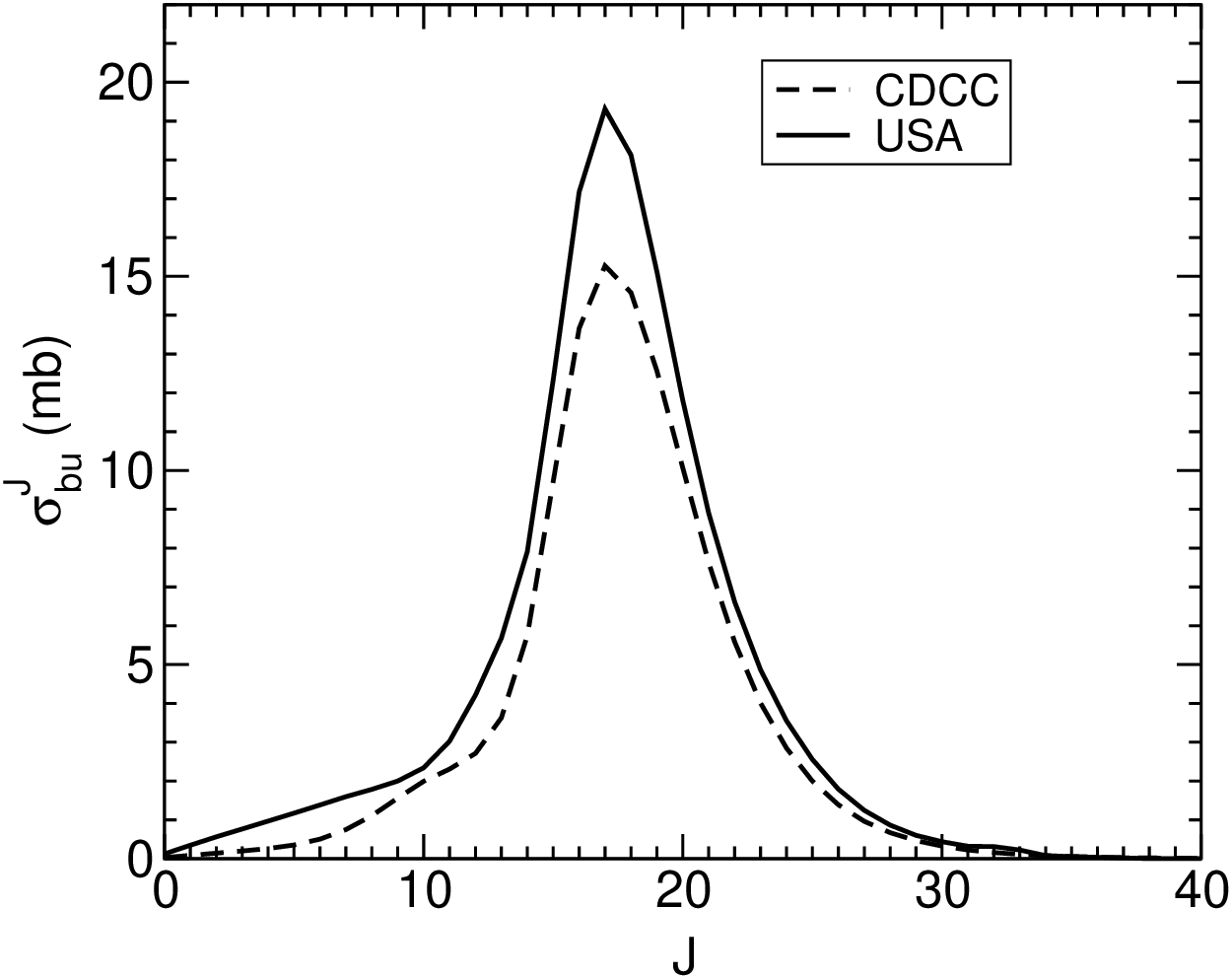}} \par}
\caption{\label{Fig:buvsJ}Integrated breakup cross section, as a function
of the total angular momentum, $J$ for the reaction
d+$^{58}$Ni at 80~MeV. The dashed and solid
line correspond to the CDCC and USA calculations, respectively. }
\end{figure}

We first study the integrated breakup
cross section, as a function of the total angular momentum $J$.
In the USA calculation, we
can make use of the closed expression (\ref{eq:xsecbuJ}). The results
are shown in Fig.~\ref{Fig:buvsJ}. It can be seen that 
the USA calculation (solid line)
reproduces the qualitative behaviour of the CDCC (dashed-line). In 
particular, both
distributions predict a maximum of the breakup cross section at $J=17$.
However, our model yields significantly more breakup than the
CDCC, as  we anticipated from the behaviour of the elastic
angular distribution.

\begin{figure}
{\centering \resizebox*{0.75\columnwidth}{!}
{\includegraphics{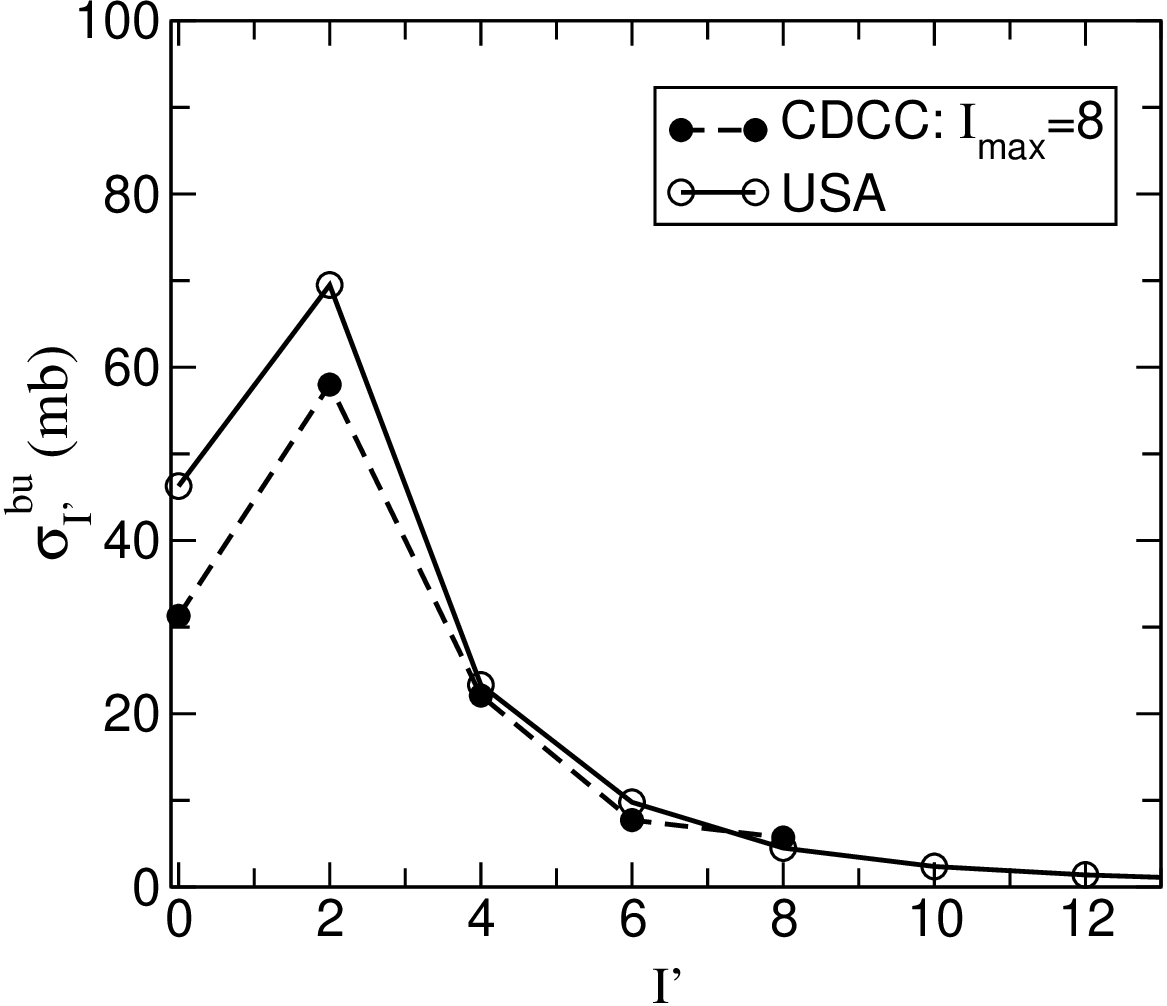}} \par}
\caption{\label{Fig:buvsI}Integrated breakup cross section, as a function
of the final p-n angular momentum, $I'$.
The filled circles (joined with dashed lines to guide the eye) correspond
to the converged CDCC calculation ($I_{\max}=8$), 
while the open circles joined by solid lines are
the USA prediction obtained with the closed expression (\ref{eq:xsecbuLIJ}). }
\end{figure}

As described in the previous section, the method here developed provides also
compact expressions for the partial integrated cross sections
leading to continuum states with definite values of \{$L'$,$I'$,$J$\},
according to Eq.~(\ref{eq:xsecbuLIJ}). Using the same
physical ingredients as in the elastic scattering, we have applied
this formula to calculate the integrated breakup cross section for $I'$=0--12.
The results are displayed by open circles in Fig.~\ref{Fig:buvsI}. It becomes
clear that the main contribution to the breakup cross sections comes from
the partial waves $I'$=0 and $I'$=2. Also included in this figure are the CDCC 
results (filled circles). Note that within CDCC  
these integrated cross sections are obtained as 
a result of large set of coupled equations,
in which all possible values of $L'$ and $I'$ need to 
be  coupled simultaneously. From this figure it becomes 
apparent that the discrepancy of the
breakup cross section between the two methods is mainly due to the
components $I'$=0 and $I'$=2. For $I'>2$ the two methods
are in excellent agreement. This result becomes more notable if one compares
the simplicity of Eq.~(\ref{eq:xsecbuLIJ}) with the large set of coupled-channels
equations  involved in the CDCC. Furthermore, 
when high excitation energies and angular momenta 
are to be included, convergence
problems typically arise in CDCC. 
This is in contrast with  
Eq.~(\ref{eq:xsecbuLIJ}), which allows one to calculate 
the breakup to highly excited 
states and large values of $I'$ in a simple way 
without the numerical difficulties involved in a CDCC calculation.

Besides the integrated breakup cross sections, we can also calculate 
the breakup to specific states of the continuum.
The $S$-matrix element describing the transition from the ground state
to a continuum state with asymptotic linear momentum $p$ and angular momenta
\{$L',I',J$\} is  given by Eq.~(\ref{eq:Sbreakup}).
Two different models where used to describe the continuum states within the USA
method. In the first one, we used a discretized continuum in terms of continuum bins,
as in the CDCC calculation. Since these bins are normalizable the overlaps
$\langle \phi_f|K \rangle_{LI}$ appearing in 
Eq.~(\ref{eq:Sbreakup}) are calculated in
exactly the same way as for the ground state. In a second calculation
we used for the final states the true scattering wavefunctions, calculated 
at a given excitation energy. Note that these states are no longer 
normalizable and so special
care has to be taken in order to calculate their overlaps with the hyperspherical 
basis. Details of the evaluation of these overlaps can be found in Appendix 
\ref{sec:ap-B}. In this calculation, we found convenient, although not essential, to 
work with analytic expressions for the wavefunctions. 
In particular, for the ground state we used the  Hulth\`en wavefunction:
\begin{equation}
\label{eq:hulthen}
\phi _{0}(r)=\sqrt{\frac{\alpha \beta (\alpha +\beta )}
{2\pi (\beta -\alpha )^{2}}}\frac{e^{-\alpha r}-e^{-\beta r}}{r} ,
\end{equation}
with $\alpha =0.2317\, \mathrm{fm}^{-1}$ and $\beta =7\alpha $.
For the continuum states, we used the solution of a 
separable potential, whose ground state coincides with the Hulth\`en wave
function, thus guaranteeing the orthogonality with the 
ground state wavefunction. These are explicitly given by 
\begin{equation}
\label{Eq:phicont}
\phi ^{(-)}_{\mathbf{p}}(\mathbf{r})=
\frac{1}{(2\pi)^3}
\left\{
e^{i\mathbf{p r}}+f(p)\frac{e^{-i p r}-e^{-\beta r}}{r} 
\right\}
\, , 
\end{equation}
with \begin{equation}
f(p)=-\left[ \beta -\frac{\beta ^{2}+p^{2}}{2\beta }-
\frac{(\beta ^{2}+p^{2})^{2}}{2\beta (\alpha +\beta )^{2}}+i p\right] ^{-1},
\end{equation}
where $\alpha$ and $\beta$ are the same as for the ground state.
These wavefunctions are normalized as 
$\langle \phi _{\mathbf{p}'}|\phi _{\mathbf{p}}\rangle =
\delta (\mathbf{p}'-\mathbf{p})$. Note also that this  potential acts 
only on the $s$-waves; for $I\neq0$, the continuum sates are
simply given by planes waves.

\begin{figure}
{\centering \resizebox*{0.9\columnwidth}{!}{\includegraphics{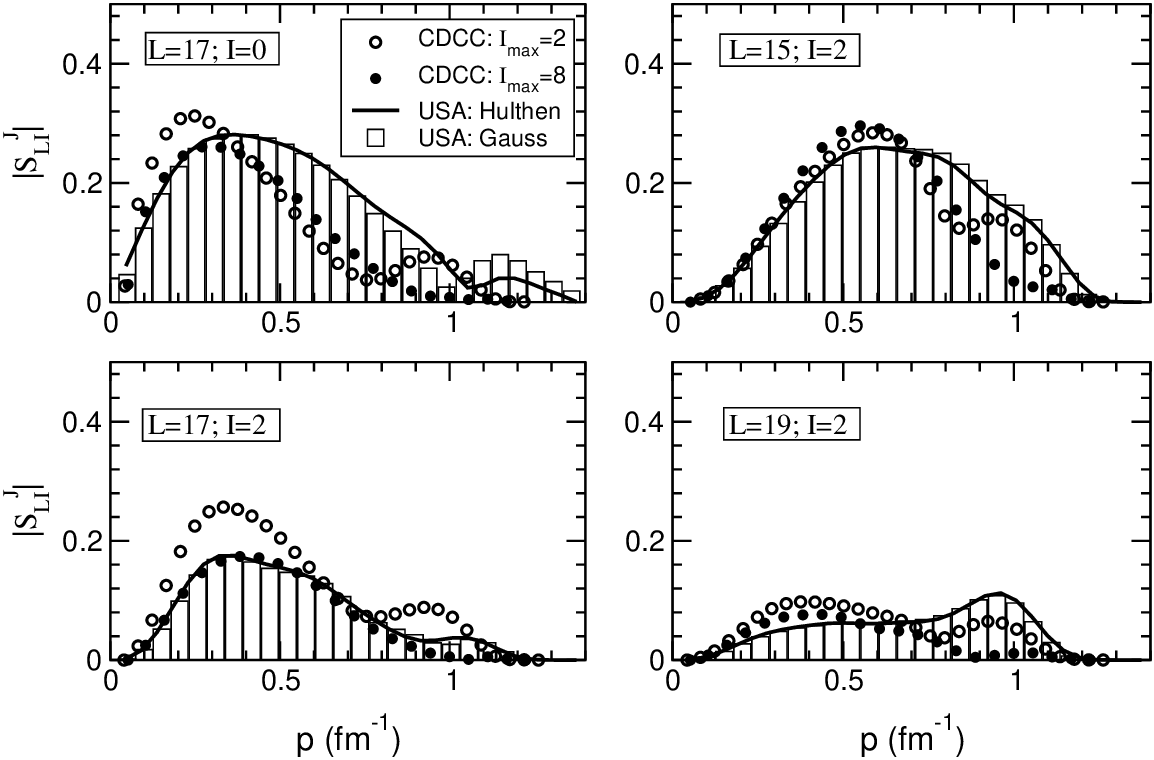}} \par}
\caption{\label{Fig:SbuJ17}Breakup \protect$S$-matrix elements
for a total angular momentum \protect$J=17\protect$ and final
p-n relative angular momentum $I'\protect$=0 and 2. The 
filled circles correspond 
to the CDCC calculation with $s$ and $d$ waves only, 
whereas the open circles are the CDCC calculations with the model space
$I=0,2,4,6,8$.
The solid line is  the USA calculation with analytic wavefunctions derived
from a separable potential.
The histogram correspond to the USA calculation using the bins 
constructed with Gaussian potential.}
\end{figure}

In Fig.~\ref{Fig:SbuJ17}, we compare the modulus of the breakup
$S$-matrix for a total angular momentum $J=17$ which, according
to Fig.~\ref{Fig:buvsJ}, corresponds to the maximum of the breakup
distribution. The continuum states with $I'=0$ and $I'=2$
have been considered for the comparison. In the latter, the separated
contributions for $L'$=15, 17 and 19 are shown. Two different CDCC
calculations are presented. The first one, represented by open circles,
uses a model space with $I=0,2$ only.
However, this model
space is not enough to achieve convergence of the $S$-matrix elements.
In analogy with \cite{Piy99}, we had to include partial waves 
up to $I_{\max}$=8. The
results of this CDCC calculation in the augmented space is shown 
by the filled circles in Fig.~\ref{Fig:SbuJ17}. 
The USA calculations with the Hulth\`en potential 
are represented in Fig.~\ref{Fig:SbuJ17} by the 
solid lines. 
Finally, the USA calculation performed with the continuum bins is 
represented with a histogram, to emphasize the fact that this calculation
uses a discretized continuum. Both calculations used a maximum hyperangular
momentum $K_m$=30. With $K_m$=40 the results are only slightly changed.
In general, we find a good
global agreement between USA and CDCC. For small values of $p$ (i.e.\
low excitation energies) the two methods yield very similar results. On the
contrary, for large values of $p$ all our calculations tend to
predict more breakup than the CDCC. Interestingly, the 
USA distributions exhibit 
a bump at $p\approx$1 fm$^{-1}$ (i.e. $\epsilon\approx 41$ MeV), which is not 
observed in the converged CDCC with $I_{\max}$=8, but appears 
nevertheless in the 
CDCC with  $I_{\max}$=2. The major disagreement between the USA and CDCC occurs
for the breakup to $s$ states. In this case, the two methods give very 
different predictions for $p>0.3$~fm$^{-1}$.

\begin{figure}
{\centering \resizebox*{0.9\columnwidth}{!}
{\includegraphics{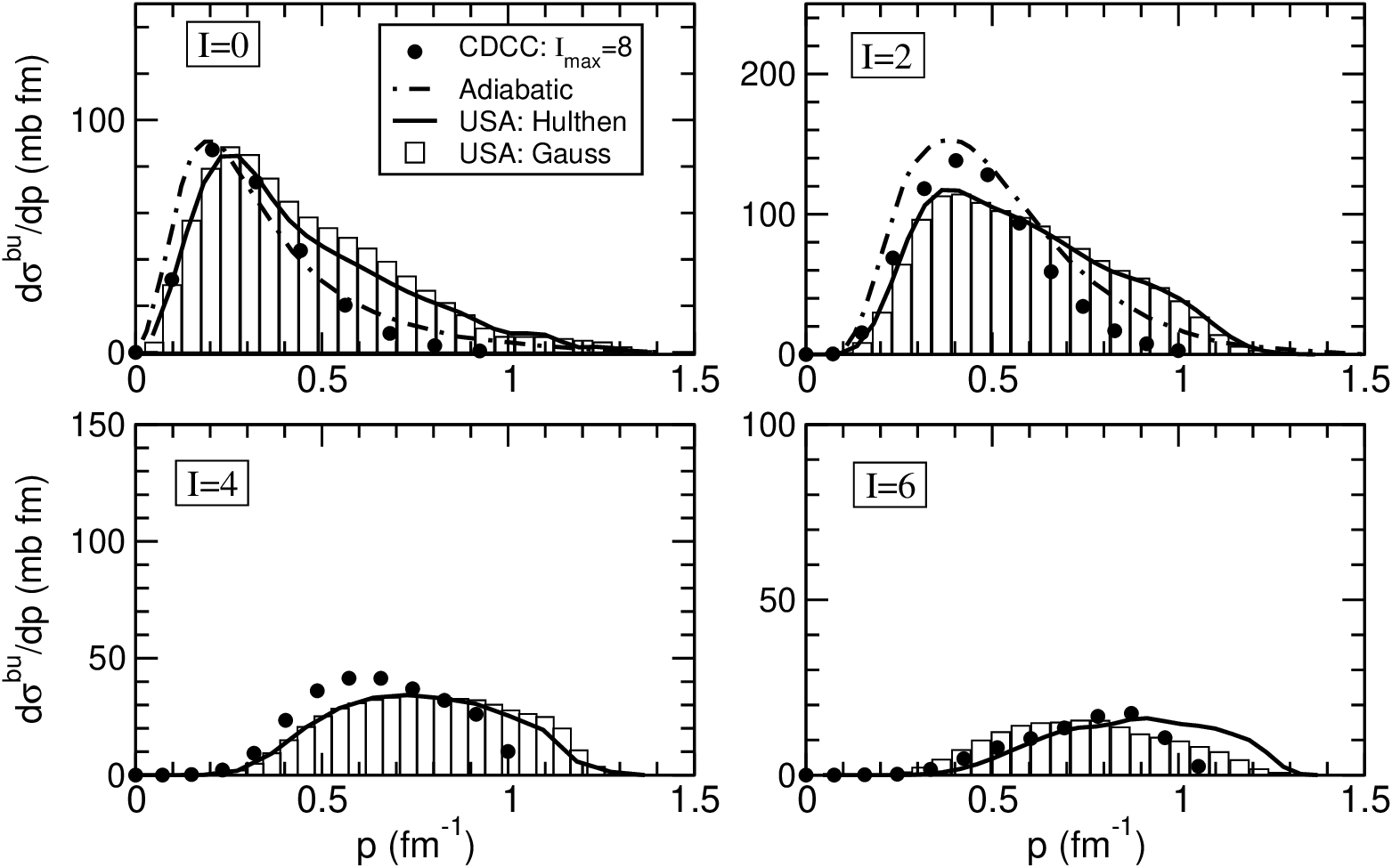}} \par}
\caption{\label{Fig:buvsq}Breakup excitation function for the reaction 
d+$^{58}$Ni at $E_d$=80~MeV. The filled circles represent 
the CDCC calculations. The
solid lines are  the USA calculations with the Hulth\`en wavefunctions. 
The histogram
is the USA calculation obtained with a Gaussian potential and a discretized continuum.
The dotted-dashed lines are the
adiabatic calculation of Amakawa and Tamura \cite{Ama82}.}
\end{figure}

%
The differential breakup cross section is readily calculated from
the above $S$-matrix elements using Eq.~(\ref{eq:difbu}), and summing
upon $L'$ and $J$. The results are displayed in 
Fig.~\ref{Fig:buvsq}. The filled
circles correspond to the CDCC calculation in the model 
space with $I_{\max}$=8.
The solid line is the USA calculation with the p-n separable interaction.
The histogram corresponds to the
USA calculation in the discretized continuum, generated with the  Gaussian
potential.
We see that, at small excitation energies, the USA and CDCC calculations
are in very good agreement. At higher excitation energies, the
USA predicts systematically more breakup, as expected from the behaviour of
the $S$-matrices analysed above. For $I'=6$, the USA calculations 
obtained with 
the continuum bins agree nicely with the CDCC.
In the cases $I'$=0 and $I'$=2 we show 
also the adiabatic coupled--channels calculation 
by  Amakawa and Tamura \cite{Ama82}. This 
calculation follows a similar trend to that of the CDCC, which is understood by
recalling that the adiabatic approximation can be regarded to an approximated CDCC 
calculation in which
the internal Hamiltonian is replaced by a constant \cite{Aus87}.

From the analysis performed in this section, we conclude that, despite
its formal simplicity, the  model developed in this work provides 
a good description of the reaction observables. In particular, we have shown
that the method describes fairly well the elastic and breakup scattering of
d+\nuc{58}{Ni} at $E_d$=80 MeV. Our comparison with the CDCC calculations
suggests that the USA tends to overestimate breakup to highly excited states
on the continuum. 

In comparing  the CDCC calculation with the USA results,  it should be taken
into account that the USA approach is not simply an approximation to the CDCC
calculation. The latter is performed assuming that the interactions 
between fragments and target, that are complex, can be approximated by local
potentials. This determines the off-shell behaviour of the interaction and, in
particular, prevents coupling to highly excited states of the projectile. 
In contrast, the USA approximation relies on the fact that off-shell matrix 
elements of the interaction are substituted by on-shell ones. This seems to 
be the reason behind the larger breakup cross sections to highly excited 
states.


\section{Summary and conclusions}
In this paper we have revisited the uncorrelated scattering approximation
(USA) originally introduced in Ref.~\cite{Mor01a}. We reviewed the basic
assumptions involved within the USA approach and analyzed its
capability to describe elastic and breakup scattering reactions. 
In what follows we summarize the main ingredients and 
results obtained in this work.

The description of the scattering reaction mechanism provided by the USA
model is based on three basic approximations.
First, in the case of a weakly bound system interacting with a heavy target,
we ignore the interaction between the fragments in the three-body
$T$-matrix propagator.
Thus, the operator $T(E)$ may be replaced by $\bar T(E) /\lambda$, being 
$\bar T(E)$ the uncorrelated $T$-matrix for a renormalized 
interaction. This has the property to conserve the angular momentum of
each fragment of the projectile with respect to the target. However, 
it results essential to conserve the interaction in the asymptotic states, which
are eigenstates of the Hamiltonian $H_0$. Expanding these states
into eigenstates of $\bar H_0$ (Hamiltonian containing only kinetic energy terms),
one finally ends up with the evaluation of the off--shell $\bar T(E)$ matrix 
elements.
 
The second approximation involved is to replace the off--shell matrix elements of
$\bar T(E)$ by the on-shell ones. This is justified because the range
of off-shellness in the matrix element is small compared with the energy
range of the operator. However, a direct substitution of the off shell
matrix element by on-shell ones may lead to the breakdown of
the unitarity property. Thus, in this work we have made use of 
a renormalization operator, which is formally equivalent to the
Democratic Mapping procedure described in \cite{Sko90}. The matrix elements of
$\bar T(E)$ between eigenstates of $\bar H_0$ can be evaluated using an 
expansion in hyperspherical harmonics. This allows us to transform analytically 
the asymptotic states into states with definite values of the angular 
momenta $L_A$ and $L_B$ of each fragment with respect to the target, and
take advantage of the fact that these magnitudes are conserved by the interaction.

The third approximation is to expand up to third
order the three-body operator
$\bar T(E)$ in terms of the three-body operators $T_A(E)$ and $T_B(E)$ 
which contain only the interaction of one of the particles with the target.
Then the on-shell matrix elements of $\bar T(E)$ can be evaluated, and the
result shows that the operator $\bar T(E)$ does not connect states in which
the total energy $E$ is distributed differently between particles $A$ and $B$.
Moreover, the matrix element of $\bar T(E)$ is completely determined by the
on-shell matrix elements of the two-body operators 
${\mathbf t}_A(e_A)$ and ${\mathbf t}_B(e_B)$
evaluated at the corresponding energies of the two particles.

Our analysis shows that the
$S$-matrix describing the scattering of a composite system by a target, 
in the uncorrelated scattering 
approximation, is given as a combination 
of products of the $S$-matrices describing
the scattering of the fragments evaluated at the corresponding energies and
angular momenta. 
The renormalization factor $\lambda$ can be obtained for each
$J$ value to ensure that, in the weak coupling limit, the USA reproduces
the elastic scattering calculated with the  folding interaction.
This renormalization factor allows us to include the effect of excitation of the
target and/or the fragments of the projectile, which is essential in nuclear 
collisions, in a way which is fully consistent with a complex folding 
interaction.

The USA keeps also some resemblance with other impulse 
approximations recently applied
to the scattering of weakly bound nuclei \cite{Cres99,Cres01a}. In analogy with
the  multiple scattering of the $T$-matrix method presented 
in \cite{Cres01a}, we start with an approximated expansion 
of the few--body $T$-matrix,
in which the inter-cluster interaction is neglected in the 
propagator. However, while in  \cite{Cres01a}
the derivation is performed in a linear momentum representation, we 
choose a partial wave description.
Moreover, although both methods express the total $T$-matrix in terms 
of two-body amplitudes, the approximations
that lead to the two--body $T$-matrices are quite different.

There exists also some formal analogy between our main 
result, Eq.~(\ref{norts}), and the semiclassical Glauber 
approximation \cite{Glau59}, 
in the sense that in both methods the scattering amplitude 
depends on a superposition of the product of 
the individual $S$-matrices of the fragments. However, despite this apparent 
similitude, we stress that the approximations
involved in both approaches are very different. First, unlike the Glauber model,
the USA does not make any 
semiclassical assumption. 
Furthermore, the Glauber model is based on the {\it frozen halo} or 
adiabatic approximation (i.e., it neglects
the excitation energies of the internal Hamiltonian), while the USA 
neglects the inter-cluster potential (the so called impulse approximation), 
but retains the full kinetic energy operator.


Finally, it should be stressed that the main differences found between the USA
and the CDCC calculations arise from the fact that the USA 
calculation gives rise 
to larger breakup cross sections to states with high excitation energies.
This result might be related to limitations of the USA approach, but it could 
also be due to the fact that the CDCC calculations assume local interactions
between the fragments of the projectile and the target, and this determines the
off-shell nature of the interactions. Accurate experimental measurements of
breakup cross sections at high excitation energies would surely help to
draw more definite conclusions on the validity of the
local, momentum independent, complex interactions used in the CDCC approach or,
by contrast, on the reliability of the presence of
relevant off-shell components in the interaction as suggested 
by the USA calculations.

\appendix
\section{\label{sec:ap-B}Expansion of the channel wavefunctions in
terms of the hyperangle}

The bound states of the projectile, as well as the normalizable bins of
continuum states, can be expressed in momentum representation
as
\begin{equation}
|\phi_i;IM_I\rangle = \int dq \phi_i(q)  |q;I M_I\rangle \, ,
\end{equation}
where the state $|q;I M_I\rangle$ is normalized so 
that $\langle q';I M_I|q;I M_I\rangle = \delta(q-q')$.
The corresponding channel states can be written as
\begin{equation}
|\phi_i,k_i;I,L,J\rangle = \int dq \phi_i(q) |q,k_i;I,L,J\rangle \, .
\end{equation}
It is convenient to write this state in terms of states with given hyperangle
$\alpha$ and kinetic energy $E_i(\alpha)$, which are given by
\begin{equation}
\tan(\alpha) = {q \over k_i}\sqrt{M \over \mu}, 
\quad E_i(\alpha) = {\hbar^2 q^2 \over 2 \mu} + {\hbar^2 k_i^2 \over 2 M} \, ,
\end{equation}
which leads to 
\begin{eqnarray}
|\phi_i,k_i;L;I,J\rangle &=& \int d\alpha g_i(\alpha)
|\alpha,E(i,\alpha);L;I,J\rangle \\
g_i(\alpha) &=&\phi_i(q(k_i,\alpha))
\left( { d q(k_i,\alpha) \over d\alpha} \right)^{1/2} \, .
\end{eqnarray}
The last term arises from the normalization 
$\langle\alpha', E|\alpha, E\rangle = \delta(\alpha'-\alpha)$.

The continuum states with asymptotic momentum $p$ are
given by
\begin{equation}
|\phi(p),k(p);I,L,J\rangle = |p,k(p);I,L,J\rangle + 
\int dq g^(-)(p,q) |q,k(p);I,L,J\rangle \, ,
\end{equation}
where the first term in the RHS represents the plane wave and the second term
represents incoming waves. The integral has a pole at $q=p$, with a residue
$R(p)$ and a 
principal part, so 
\begin{eqnarray}
 |\phi(p),k(p);I,L,J\rangle & =&  [1+ i \pi R(p)] |p,k(p);I,L,J \rangle 
\nonumber \\
& + & 
{\mathcal P}\int dq g^(-)(p,q)  |q,k(p);I,L,J\rangle \, .
\end{eqnarray}
This state can be written in terms of the hyperangle as
\begin{eqnarray}
|\phi(p),k(p);I,L,J\rangle &=& [1+ i \pi R(p)] 
\left({ d p \over d\alpha_p} \right)^{1/2}|\alpha_p,E;I,L,J\rangle 
\nonumber \\
 &+& {\mathcal P} \int d \alpha g^(-)(p,q(\alpha)) 
\left({ d q \over d\alpha} \right)^{1/2} |\alpha,E(p,\alpha);I,L,J\rangle \, ,
\nonumber \\
\end{eqnarray}
where
\begin{eqnarray}
\tan(\alpha_p) = {p \over k(p)}\sqrt{M \over \mu}, &\quad&
E={\hbar^2 p^2 \over 2 \mu} + {\hbar^2 k(p)^2 \over 2 M} \, ,\\
\tan(\alpha) = {q \over k(p)}\sqrt{M \over \mu}, &\quad&
 \quad E(p,\alpha) = {\hbar^2 q^2 \over 2 \mu} + {\hbar^2 k(p)^2 \over 2 M} \, .
\end{eqnarray}

\section{\label{sec:ap-A}On-shell matrix elements in the hyperspherical basis}
In this appendix we show in detail the procedure used to evaluate the on-shell
matrix elements of the operator $\bar T(E)$. First, we make use of an
expansion in hyperspherical harmonics which allows us to write down
\begin{equation}
\int d\alpha  g^{I}_i(\alpha) A^i(\alpha) |\alpha,E;L,I,J\rangle 
= \sum_K \langle K|\phi_i\rangle_{LI} |K,E;L,I,J\rangle\, ,
\end{equation}
where we introduce the states
\begin{equation}
|K,E;L,I,J\rangle  = \int_{0}^{\pi/2} d\alpha f^K_{L,I}(\alpha) |\alpha,E;L,I,J\rangle .
\end{equation}
The functions $f^K_{L,I}(\alpha)$ are given by
\begin{equation}
f_{L,I}^{K}(\alpha ) = N^{LI}_{K}(\cos \alpha )^{L+1}
(\sin \alpha )^{I+1}P^{(I+\frac{1}{2},L+\frac{1}{2})}_{(K-L-I)/2}(\cos 2\alpha ), 
\label{eq:f(I,L,K)}
\end{equation}
where $P^{(I+\frac{1}{2},L+\frac{1}{2})}_{n}$ are the Jacobi polynomials
of degree $n$
and $N^{LI}_{K}$ are some normalization constants, 
whose explicit expressions can 
be found in \cite{Mor01a}.

The coefficients  $\langle K|\phi_i\rangle_{LI}$ are explicitly given by
\begin{eqnarray}
\langle K|\phi_i\rangle_{LI} &=& \sum_{K'} c^i_{LI}(K') A^i_{LI}(K',K)  \\
c^i_{LI}(K') &=& \int d \alpha f^{K'}_{L,I}(\alpha)g^{I}_i(\alpha) \\
A^i_{LI}(K',K) &=& \int d \alpha f^{K'}_{L,I}(\alpha) A^i(\alpha)
f^{K}_{L,I}(\alpha) \, .
\end{eqnarray}
To get these expressions we have made use of the closure property of the 
hyperspherical harmonics
\begin{equation}
\sum_K f^{K}_{L,I}(\alpha) f^{K}_{L,I}(\alpha') = \delta(\alpha-\alpha') \, .
\end{equation}
Then, we end up with the following expression for the $T$-matrix elements
\begin{eqnarray}
&& \langle\phi_f,k_f;L',I',J| T(E)|\phi_g,k_g;L,I,J\rangle  
\simeq  \nonumber \\
&& \sum_{K',K} \langle\phi_f|K'\rangle_{L'I'} \langle K|\phi_g\rangle_{LI}  
 \langle K',E;L',I',J|\bar T(E)|K,E;L,I,J\rangle \, .
\label{tsumk}
\end{eqnarray}
The requirement that the approximated expression preserves unitarity for 
hermitian interactions leads to
\begin{equation}
\sum_i \langle K'|\phi_i\rangle_{LI} \langle\phi_i|K\rangle_{LI} =  \delta_{K,K'} \, .
\label{closure}
\end{equation}
On the other hand, if the interaction is constant, the matrix elements
of $\bar T(E)$ will be diagonal and independent on $K$. In this case,
the $T$-matrix should not couple different internal states. 
This leads to
\begin{equation}
\sum_K \langle K|\phi_g\rangle_{LI} \langle\phi_f|K\rangle_{LI} =  \delta_{g,f} \, .
\end{equation}
These conditions can be achieved by an adequate choice of $A^i_{LI}(K',K)$. 
In what follows we shall assume, for definiteness, that in each spin channel $I$ there is
at most one bound state, while the rest of the states correspond to the 
continuum.
So, for the bound state in channel $I$ we take 
\begin{eqnarray}
A^b_{LI}(K,K') &=& \delta_{K,K'} /\sqrt{P^b_{LI}}\, ,
\end{eqnarray}
with $P^b_{LI} = \sum_K |c^b_{LI}(K)|^2$,
while for the rest of the states in channel $I$ we consider a unique symmetric 
matrix $A^c_{LI}(K',K)$ which fulfils
\begin{eqnarray}
&& \sum_{K'} A^c_{LI}(K,K') c^b_{LI}(K') = 0 \, ,  \\
&&  \sum_{K_1,K'_1} A^c_{LI}(K,K_1)A^c_{L,I}(K',K'_1) 
\sum_{i\ne b} c^{i*}_{LI}(K_1) c^i_{LI}(K'_1) 
= \delta_{K,K'} - \frac{c^b_{LI}(K)c^b_{LI}(K')}{P^b_{LI}} \, .\nonumber \\
&&
\end{eqnarray}
Note that this procedure is equivalent to orthogonalize all the continuum 
states with respect to the ground state,  and then apply the ``democratic
mapping'' procedure to the continuum states.

It should be noticed that the sum in $K$ is extended in principle up to
infinity. Note that in this case  $P^b_{L,I}=1$. In practice, the sum is 
taken up to a maximum value $K_m$, which is
obtained to get convergence in the calculations. So, in general, 
$P^b_{L,I} \le 1$,
specially for the higher partial waves, for which $L$ is close to $K_m$.
Note that, in contrast to what was done in 
Ref.~\cite{Mor01a,Mor01b}, no physical meaning is attached to the parameter $K_m$, 
which should be taken as large as possible, until convergence is achieved.

In order to make use of the fact that $\bar T(E)$ conserves the angular 
momenta of the fragments, one can use the 
Raynal-Revai transformation \cite{Ray70}:
\begin{equation}
|K,E;L,I,J\rangle = \sum_{L_A,L_B} |K,E;L_A,L_B,J\rangle\langle L_A,L_B|L,I\rangle_{KJ} \, .
\end{equation}
The states with different $K$ values can be expanded to get
states with definite values of the kinetic energy of each fragment. 
We can write, in terms of the hyperangle,
\begin{equation}
|K,E;L_A,L_B,J\rangle = \int d \beta f^K_{L_A,L_B}(\beta) |\beta,E;L_A,L_B,J\rangle \, .
\label{app1}
\end{equation}
This state can be characterized in terms of a product state of particles
$A$ and $B$ with energies $e_A=E \cos^2(\beta)$ and $e_B= E \sin^2(\beta)$.


\section{\label{sec:ap-C} Multiple scattering expansion of the $\bar T$ 
operator}
In this appendix we evaluate the on-shell matrix elements  
$\langle\beta',E|\bar T(E)|\beta,E\rangle$, where the state $|\beta,E\rangle$
is an eigenstate of $\bar H_0 = K_A + K_B$, corresponding to energies 
$e_A=E \cos^2 \beta$ and $e_B=E \sin^2 \beta$. Analogously, the state
$|\beta',E\rangle$ corresponds to $e'_A=E \cos^2 \beta'$ and $e'_B=E \sin^2 \beta'$. Note that $E=e_A+e_B=e'_A+e'_B$.
We can write explicitly
\begin{equation}
|\beta, E\rangle = N(\beta) |e_A, e_B\rangle \, ,
\end{equation}
where $N(\beta)= \sqrt{E \sin (2 \beta)}$ is  the 
square root of  the Jacobian of the transformation from
$\{e_A, e_B\}$ to $\{\beta, E\}$.

We consider
the multiple scattering expansion of the operator $\bar T(E)$ up to  third 
order
\begin{eqnarray}
\bar T(E) & = & 
T_A(E) + T_B(E) \nonumber \\ 
\label{eq:TAB}
& + & T_A(E) \bar G_0(E) T_B(E) + T_B(E) \bar G_0(E) T_A(E) \nonumber \\
&+&
T_A(E) \bar G_0(E) T_B(E) \bar G_0(E) T_A(E) + 
T_B(E) \bar G_0(E) T_A(E) \bar G_0(E) T_B(E)\, ,  \nonumber \\
%
\end{eqnarray}
where $\bar G_0(E)=(E^+-K_A-K_B)^{-1}$ is the free 3-body propagator and 
\begin{equation}
T_A(E) = \lambda v_{AT} + \lambda v_{AT} \bar G_0(E) T_A(E) \, .
\label{eq:TA}
\end{equation}
The matrix operator $T_A(E)$ can change the kinetic energy
of particle $A$, through its off--shell components, but it can not modify
the kinetic energy of $B$. Hence, we can write
\begin{equation}
\langle\beta',E|T_A(E)|\beta,E\rangle  = 
\delta(\beta-\beta')\langle e_A|{\bf t}_A(e_A)|e_A\rangle \, , \\
\end{equation}
where we have introduced the operator  
\begin{equation} 
{\bf t}_A(e_A)= \lambda v_{AT} +  \lambda v_{AT}  g_A(e_A){\bf t}_A(E) ,
\label{eq:tA}
\end{equation} 
with $g_A(e)=(e^+-K_A)$ the free 2-body propagator.
Note that the operators $T_A$ and ${\bf t}_A$ differ on the kinetic operator appearing
in the propagator. While $T_A$ is defined with the full kinetic energy
operator, i.e., $K_A+ K_B$, the propagator in ${\bf t}_A$ contains  only 
the kinetic energy operator associated
with particle $A$, $K_A$. Therefore, $T_A$ should be 
understood as a three-body operator,
whereas ${\bf t}_A$ corresponds to  a two-body operator. For simplicity,
in the following, we drop the energy argument of the three-body operators $T_A, T_B,
\bar G_0$, which is in all cases $E$, 
but we retain it in the two-body operators.

%
%
The contribution of the second order terms can be expressed as the sum of a 
pole term and a principal part
\begin{eqnarray}
\label{second1}
\langle\beta',E| T_A \bar G_0 T_B |\beta,E\rangle \nonumber
&= & i \pi \delta(\beta-\beta') t_A(e_A)t_B(e_B) \\
& + & {\langle e_A'|{\bf t}_A(e'_A)|e_A\rangle
\langle e'_B|{\bf t}_B(e_B)|e_B\rangle\over E-e'_B-e_A} 
N(\beta) N(\beta')\, , \\
\label{second2}
\langle\beta',E| T_B \bar G_0 T_A |\beta,E\rangle \nonumber
&= & i \pi \delta(\beta-\beta') t_A(e_A)t_B(e_B) \\
& + & {\langle e'_B|{\bf t}_B(e'_B)|e_B\rangle
\langle e'_A|{\bf t}_A(e_A)|e_A\rangle\over E-e'_A-e_B}N(\beta) N(\beta') \, , \
\end{eqnarray}
where we have introduced the short notation $t_A(e_A)$ and $t_B(e_B)$ 
for the on-shell matrix elements
$\langle e_A|{\bf t}_A(e_A)|e_A\rangle$ and 
$\langle e_B|{\bf t}_B(e_B)|e_B\rangle$, 
respectively.

%
%
As for the third order terms we have
\begin{eqnarray}
&& 
\langle\beta',E | T_A \bar G_0 T_B \bar G_0 T_A | \beta,E\rangle 
\nonumber \\
&=&
\int de''_A \langle e'_A | t_A(e'_A) | e''_A \rangle 
\frac{1}{E^+-e''_A-e'_B} 
\langle e'_B | t_B(E-e''_A) | e_B \rangle 
\nonumber \\
&\times & \frac{1}{E^+-e''_A-e_B} 
\langle e''_A | t_A(e_A) | e_A \rangle N(\beta) N(\beta') \, .
\end{eqnarray}
To evaluate this contribution we make use of the following identities
\begin{eqnarray}
g_A(e')g_A(e)&=& \frac{1}{e-e'} \Big(g_A(e')-g_A(e)\Big)\\
t_A(e')-t_A(e) &=& t_A(e') \Big(g_A(e')-g_A(e)\Big) t_A(e) \, .
\label{toff}
\end{eqnarray}
The second of these expressions indicates that the operator $ t_B(E-e''_A)$
can be approximated by $t_B(e_B)$, and the difference would be of higher order
in the $T$-matrix expansion. Thus, one may write
\begin{eqnarray}
&& 
\langle\beta',E | T_A \bar G_0 T_B \bar G_0 T_A | \beta,E\rangle 
\nonumber \\
&&
\simeq
\langle e'_B | {\bf t}_B(e_B)| e_B\rangle 
\langle e'_A | {\bf t}_A(e'_A) g_A (e'_A) g_A (e_A) {\bf t}_A(e_A)| e_A \rangle 
N(\beta) N(\beta') \, .
\end{eqnarray}

Making use of the above identities, one gets
\begin{eqnarray}
\label{third1}
&& 
\langle\beta',E | T_A \bar G_0 T_B \bar G_0 T_A | \beta,E\rangle 
\nonumber \\
& \simeq & \langle e'_B | {\bf t}_B(e_B)| e_B\rangle 
\frac{\langle e'_A | {\bf t}_A(e'_A) | e_A\rangle - 
\langle  e'_A |{\bf t}_A(e_A)| e_A \rangle }{e_A-e'_A}
 N(\beta) N(\beta') \, .
\end{eqnarray}
A similar derivation for the other third order term gives
\begin{eqnarray}
\label{third2}
&& 
\langle\beta',E | T_B \bar G_0 T_A \bar G_0 T_B | \beta,E\rangle 
\nonumber \\
& \simeq &
 \langle e'_A | {\bf t}_A(e_A)| e_A\rangle 
\frac{ \langle e'_B | {\bf t}_B(e'_B) | e_B\rangle - 
\langle  e'_B |{\bf t}_B(e_B)| e_B \rangle  }{e_B-e'_B}
N(\beta) N(\beta') \, .
\end{eqnarray}
So, the sum of the two third order contributions reduces to
\begin{eqnarray}
\label{third}
&& \langle\beta',E | T_A \bar G_0 T_B \bar G_0 T_A | \beta,E\rangle 
+
\langle\beta',E | T_B \bar G_0 T_A \bar G_0 T_B | \beta,E\rangle 
 \nonumber \\
&& =
\frac{\langle e'_B|{\bf t}_B(e_B)|e_B\rangle \langle e'_A|{\bf t}_A(e'_A)|e_A\rangle }{e_A-e'_A} N(\beta) N(\beta')
 \nonumber \\
&& +
\frac{\langle e'_A|{\bf t}_A(e_A)|e_A\rangle \langle e'_B|{\bf t}_B(e'_B)|e_B\rangle 
}{e_B-e'_B}N(\beta) N(\beta') \, ,
\end{eqnarray}
which cancels exactly the principal value part of the second order terms, 
Eqs.~(\ref{second1}) and (\ref{second2}). Therefore, collecting these results, we
have that the on-shell matrix elements of the $\bar T$ operator up to third order
can be written as
\begin{eqnarray}
\langle\beta',E| \bar T(E) |\beta,E\rangle
=  \delta(\beta-\beta')
\Big\{  t_A(e_A) + t_B(e_B) + 2 i \pi  t_A(e_A)t_B(e_B) \Big\} \, .
\nonumber \\
\end{eqnarray}

%
%
\ack We acknowledge fruitful discussions with 
R. Crespo, J. Raynal and R. C. Johnson. This work has been 
partially supported by the 
Spanish MCyT projects FPA2002-04181-C04-04 and 
BFM2002-03315. A.M.M. acknowledges 
a postdoctoral grant from 
the Funda\c c\~ao para a Ciencia e a Tecnologia (Portugal).

\bibliographystyle{elsart-num}
\bibliography{./usa}
\end{document}